\documentclass[aps,prl,reprint,groupedaddress,showpacs,graphicx]{revtex4-1}
\pdfoutput=1

\usepackage{amsmath,amssymb,amstext,bm}
\usepackage[pdftex]{graphicx}
\usepackage{siunitx}
\usepackage{physics}
\usepackage{subcaption}
\graphicspath{{images/}}

\usepackage{hyperref}
\usepackage{breakurl}

\begin{document}

\title{Chiral symmetry-breaking dynamics in the phase transformation of nematic droplets}

\author{Fred Fu}
\author{Nasser Mohieddin Abukhdeir}
\altaffiliation{Department of Physics and Astronomy, University of Waterloo, Waterloo, Ontario, Canada}\
\altaffiliation{Waterloo Institute for Nanotechnology, University of Waterloo, Waterloo, Ontario, Canada}\
\email[E-mail: ]{nmabukhdeir@uwaterloo.ca}
\homepage[Homepage: ]{http://chemeng.uwaterloo.ca/abukhdeir/}

\affiliation{Department of Chemical Engineering, University of Waterloo, Waterloo, Ontario, Canada}

\date{\today}

\begin{abstract} 
	Dynamic simulations of the isotropic-nematic phase transformation of liquid crystal droplets under homeotropic anchoring are found to predict chiral symmetry-breaking dynamics.
	These observations occur when using material parameters for pentyl-cyanobiphenyl (5CB) but not under the single elastic constant approximation of this material, frequently used in simulation.
	The twisting dynamic process occurs during the relaxation of the domain from an unstable radial texture to a stable uniform texture and involves simultaneous defect loop motion and twisting of the bulk nematic texture.
\end{abstract}

\maketitle


Chiral symmetry-breaking equilibrium textures in deformable and nondeformable droplets of achiral nematic liquid crystals (LC) have long been observed for a broad range of LC compounds and conditions \cite{Volovik1983,Williams1986a,Drzaic1999,Ohzono2012,Vanzo2012}.
Press and Arrott \cite{Press1974} first predicted equilibrium twisted radial textures in nematic droplets and found them stable for nematic LCs with twist deformations preferred over splay and bend.
Since then, the existence of a wide variety of twisted equilibrium textures have been observed both experimentally \cite{Volovik1983,Drzaic1999,Ohzono2012} and via simulation \cite{Williams1986a,Vanzo2012} for nematic domains with length scales ranging from nanoscale to macroscale.
Recently, chiral symmetry breaking in nematics has taken on even more importance due to the discovery of chromonic and bent-core nematic compounds which exhibit previously unobserved nematic elasticity properties \cite{Jeong2014}.

Most of the observed chiral symmetry-breaking phenomena in achiral nematic LCs have involved spherical nematic droplets, which serve as ideal geometries for studying the interplay between surface and bulk elastic effects on nematic texture.
Furthermore, the study of nematic droplets is technologically relevant to applications such as biological sensing \cite{Lin2011} and polymer-dispersed liquid crystal (PDLC) film-based smart glass \cite{Drzaic1995}.

Experimental research on nematic LC domains has been limited to relatively large length ($\si{\micro\meter}$) and time ($\si{\milli\second}$) scales compared to the characteristic scales of nematic LCs ($\si{\nano\meter}$, $\si{\micro\second}$).
Nematic textures are typically analyzed using polarized optical microscopy or fluorescence confocal polarizing microscopy \cite{Lavrentovich2003}, which has adequate resolution for LC domains of at least a few micrometers in diameter.
While deuterium nuclear magnetic resonance \cite{Crawford1992} and other more advanced imaging methods \cite{Higgins2005} are able to access dynamics at \si{\micro\meter} and \si{\micro\second} resolutions, it still remains highly nontrivial to study texture dynamics for nanoscopic droplets at relevant time scales.
Alternatively, simulations using molecular and continuum models have shown significant promise for augmenting experimental research through direct access to characteristic LC scales.
Continuum simulations, recently reviewed in ref. \cite{Abukhdeir2016a}, have been increasingly able to capture complex LC physics including phase formation, defect dynamics, heat transfer, and hydrodynamics.

In this work, continuum simulations are performed of the formation of an initially isotropic phase nematic LC confined within a nanoscale nondeformable droplet with weak homeotropic surface anchoring conditions.
Past research has shown that as the size of a nematic droplet decreases, the equilibrium nematic texture transitions from a radial configuration with a $+1/2$ disclination loop to a defect-free uniform texture \cite{Lavrentovich1998}.
This transition results from the competition between the bulk elasticity and surface anchoring effects.
The simulations presented here are performed within the droplet size and surface anchoring strength regime where the uniform texture is stable.
Beginning at the initial quench, nematic formation occurs, initially forming an unstable radial-like texture.
Subsequently, a spontaneous symmetry-breaking twist-mediated defect escape mechanism is observed for material parameters corresponding to pentyl-cyanobiphenyl (5CB) prior to the droplet evolving to the equilibrium uniform texture.
Notably, this dynamic mechanism does not occur under the single-constant approximation frequently used in LC simulations.
In such an approximation, physically realistic differences between the modes of elastic deformation are neglected and the elastic free energy is parameterized by a single elastic constant.

Nematic order is modelled through the use of a symmetric-traceless second-rank tensor order parameter or alignment tensor \cite{deGennes1995,Sonnet1995}
\begin{equation}\label{eqn:qtensor}
    Q_{ij} = S\left(n_in_j - \frac{1}{3}\delta_{ij}\right) + P(m_i m_j - l_i l_j)
\end{equation}
where $S \in [-0.5,1]$ quantifies the degree of local uniaxial alignment, $\bm{n}$ is the nematic director (major eigenvector), and $P$ quantifies the degree of biaxial alignment associated with the two minor eigenvectors $\bm{m}$ and $\bm{l}$.
The Landau--de Gennes (LdG) free energy density is given by \cite{deGennes1995}
\begin{equation}
    f - f_{iso} = f_b(\bm{Q}, T) + f_{el}(\bm{Q},\bm{\nabla Q})
\end{equation}
where the bulk or thermodynamic contribution is
\begin{equation}
    f_{b} = \frac{1}{2}a_0(T-T_{ni})(Q_{ij}Q_{ji}) + \frac{1}{3}b(Q_{ij}Q_{jk}Q_{ki}) + \frac{1}{4}c(Q_{ij}Q_{ji})^2
\end{equation}
where $a_0$, $b$, and $c$ are material parameters and $T_{ni}$ is the theoretical second-order nematic-isotropic transition temperature.
The elastic contribution to the free energy density is \cite{Barbero2001,Mori1999}
\begin{align}\label{eqn:q_elasticity}
    f_{el} =& \frac{1}{2}L_1(\partial_i Q_{jk} \partial_i Q_{kj}) + \frac{1}{2}L_2(\partial_i Q_{ij} \partial_k Q_{kj}) \nonumber\\
    &+ \frac{1}{2}L_3(Q_{ij}\partial_iQ_{kl} \partial_jQ_{kl}) + \frac{1}{2}L_{24}(\partial_kQ_{ij} \partial_jQ_{ik})\nonumber\\
    =& \frac{1}{2}L_1 G_1 + \frac{1}{2}L_2 G_2 + \frac{1}{2} L_3 G_3 + \frac{1}{2}L_{24}G_{24}
\end{align}
which incorporates the full anisotropy in nematic elasticity for nonzero elastic constants $L_i$.

The values used for the thermodynamic parameters are documented in ref. \cite{Fu2017}.
Elastic parameters $L_i$ are estimated from experimental measurements of the elastic constants from the Frank-Oseen free energy density \cite{Frank1958}:
\begin{multline}\label{eqn:frank}
f_{f} = \frac{1}{2}k_{11}(\div{\bm{n}})^{2}
+\frac{1}{2}k_{22}(\bm{n}\vdot\curl{\bm{n})^{2}}
+\frac{1}{2}k_{33}(\bm{n}\cross\curl{\bm{n}})^{2} \\ -\frac{1}{2}(k_{22}+k_{24})\div{\left((\bm{n}\cross\curl{\bm{n}})+\bm{n}(\div{\bm{n}})\right)}
\end{multline}
where coefficients $k_{11}$, $k_{22}$, $k_{33}$ and $k_{24}$ are elastic constants associated splay, twist, bend, and saddle-splay nematic deformations, respectively.
For 5CB, Frank-Oseen elastic constants were used from past work \cite{Bogi2001,Dunmur2001,Polak1994,Pairam2013} where $k_{11}$ and $k_{33}$ were take from ref.~\cite{Bogi2001}, $k_{22} = 0.66k_{11}$ from ref. \cite{Polak1994}, and $k_{24} = k_{22}$ from refs. \cite{Pairam2013,Polak1994}.

The appropriate conversion to the LdG elastic constants is then performed under the assumption of uniaxial $\bm{Q}$ and uniform $S$ \cite{Mori1999}:
\begin{align}
L_1 =& \frac{1}{6S^2}(k_{33} - k_{11} + 3k_{22})\label{eqn:L1}\\
L_2 =& \frac{1}{S^2}(k_{11} - k_{22} - k_{24})\\
L_3 =& \frac{1}{2S^3}(k_{33} - k_{11})\\
L_{24} =& \frac{1}{S^2}k_{24}.\label{eqn:L24}
\end{align}
This results in the following constants assuming $S=S_{eq}$ where $S_{eq}$ is the bulk scalar order parameter at equilibrium: $L_1 = \SI{9.1}{\pico\newton}$, $L_2 =\SI{-8.1}{\pico\newton}$, $L_3 = \SI{7.1}{\pico\newton}$, and $L_{24} = \SI{16.6}{\pico\newton}$.

Homeotropic surface anchoring, where the preferred nematic director parallel to the surface normal, is modelled using a quadratic surface free energy density
\begin{equation}
    f_s = \frac{1}{2}\alpha (Q_{ij} - Q_0)^2
\end{equation}
where $Q_0 = S_{eq}(\nu_i\nu_j - \frac{1}{3}\delta_{ij})$ and $\bm{\nu}$ is the surface unit normal \cite{Nobili1992}.
A surface anchoring strength $\alpha = \SI{e-5}{J\per m^2}$ is used, corresponding to weak anchoring \cite{Lavrentovich1998,Drzaic1995}.
At the simulated temperature $T=\SI{307}{K}$, which is slightly below $T_{ni}$ for 5CB, $S_{eq} \approx 0.31$ \cite{Fu2017}.

Nematic dynamics are simulated using a time-dependent Ginzburg-Landau model (model A dynamics) \cite{Hohenberg1977},
\begin{equation}\label{eqn:governing}
    \frac{\partial Q_{ij}}{\partial t} = -\mu_{r}^{-1} \left[\frac{\delta F}{\delta Q_{ij}}\right]^{ST},
\end{equation}
where $F = \int_{V} f \, dV + \int_{S} f_{s} \, d S$ is the total free energy, $\mu_{r}$ is the rotational viscosity of the nematic phase, approximately $\mu_{r} = \SI{0.055}{Ns\per m^2}$ for 5CB \cite{Skarp1980}, and $[\text{ }]^{ST}$ indicates the symmetric traceless component of the tensor.
The initial condition used for all simulations is an isothermal isotropic phase droplet immediately following a quench below $T_{ni}$ with a thin homeotropically-aligned boundary layer at the droplet surface, corresponding to a heterogeneous nucleation \cite{Sheng1982}.
Numerical solution of the model was performed using the finite element method \cite{FEniCSBook} with tetrahedral mesh elements of scale of approximately $\SI{10}{nm}$, on the order of the nematic coherence length.
The droplet geometry used was slightly oblate, being compressed by $2\%$ of the sphere diameter along the $z$-axis.
This oblateness results in a small preference for alignment along the $z$ direction which eliminates a possible degeneracy in overall droplet alignment that arises from a perfectly spherical shape.
This degeneracy could result in a saddle point in the free energy leading to anomalous dynamics \cite{Khayyatzadeh2014}.

\begin{figure*}
	\includegraphics[width=\linewidth]{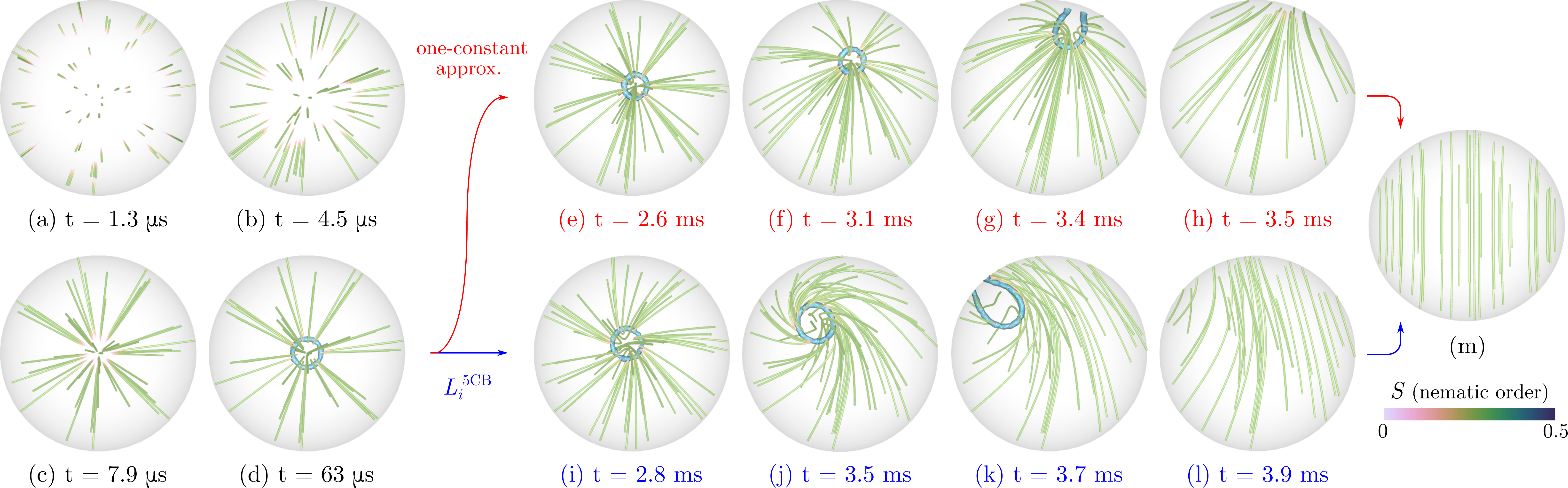}
	\caption{(Color online) Hyperstreamline visualizations of the evolution of the alignment tensor field for a droplet of $R=\SI{375}{nm}$: (a--d) Nematic phase formation (qualitatively identical for both elasticity cases), (e--h) defect loop escape (one-constant elasticity), (i--l) defect loop escape (5CB elasticity), and (m) the equilibrium uniform texture domain. For (a--l) the z-axis is oriented out-of-plane and for (m) it is reoriented in-plane vertical (for clarity).}
	\label{fig:texture-evol}
\end{figure*}

Simulations were performed using both 5CB elasticity parameters and the single elastic constant approximation in which $L_1 > 0$ and $L_2 = L_3 = L_{24} = 0$.
This corresponds to $k_{11} = k_{22} = k_{33}$, for which the experimental estimate of $k_{11}$ for 5CB is used, and $k_{24}=0$ \cite{Mori1999}.
Figure~\ref{fig:texture-evol} shows visualizations of the time evolution of the alignment tensor field $\bm{Q}$ for both elasticity cases, showing nematic phase formation into a radial nematic texture followed by relaxation to a uniform texture via disclination motion and escape.
The alignment tensor field is visualized using hyperstreamlines \cite{Delmarcelle1993,Fu2015} where their direction corresponds to the local orientation of the nematic director, color to the magnitude of $S$, and eccentricity of their cross section to the magnitude of $P$.

For both cases, the initial formation dynamics (Figure~\ref{fig:texture-evol}a--d) observed are consistent with past dynamic simulation results \cite{Fu2017}, in which the stable nematic phase initially grows freely into the center of the domain but eventually slows due to capillary effects and forms a $+1/2$ disclination loop.
The resulting fully-formed nematic domain has a radial nematic texture (Figure~\ref{fig:texture-evol}d) due to topological constraints imposed through the combination of homeotropic anchoring and spherical confinement.
However, the competition between surface anchoring and bulk elasticity results in this radial texture being unstable, causing relaxation of the nematic domain to continue toward a uniform defect-free texture (Figure~\ref{fig:texture-evol}m), which is expected for cases of small nanoscale droplets \cite{Lavrentovich1998}.

The relaxation mechanism for the single elastic constant approximation simulation, shown in Figures~\ref{fig:texture-evol}e--h, is  significantly different compared to that for the 5CB elasticity case, shown in Figures~\ref{fig:texture-evol}i--l.
For the single elastic constant case, the relaxation process occurs in an intuitive way, through simultaneous defect loop motion towards the domain boundary and bulk texture relaxation.
As the defect loop approaches the boundary, the local anchoring deviates significantly from homeotropic alignment and the defect loop ``escapes'' through the boundary.
However, for the 5CB elasticity case, as the defect loop translates toward the boundary, the alignment in the surrounding region also rotates about the axis orthogonal to the plane of the loop.
In this way, defect motion is accompanied by a simultaneous twist deformation in the bulk texture.
As the defect loop escapes the boundary through local relaxation of the homeotropic anchoring, the bulk nematic texture eventually unravels into a uniform state.
Qualitatively, this mechanism results in decreased splay deformation compared to the single-constant case, at the expense of increased twist deformation.

In order to analyze these dynamic processes quantitatively, it is useful to approximate the individual contributions to the tensorial elastic free energy density (Equation~\ref{eqn:q_elasticity}) with respect to the different canonical modes of nematic deformation from the Frank-Oseen model (Equation~\ref{eqn:frank}).
In order to estimate these contributions from the tensorial LdG form of nematic elasticity, the uniaxial component of $\bm{Q}$ was first calculated for each simulation, corresponding to the first term on the right side of Equation~\ref{eqn:qtensor}.
The derived relations between $L_i$ and $k_i$ (Equations~\ref{eqn:L1}--\ref{eqn:L24}) \cite{Mori1999} were then substituted into Equation~\ref{eqn:q_elasticity} such that $f_{el}$ is parameterized in terms of $k_i$:
\begin{align}
    f_{el} &= \frac{1}{2 S^2} k_{11} \left(-\frac{1}{6} G_1 + G_2 - \frac{1}{2 S} G_3 \right) \nonumber\\
    &+ \frac{1}{2 S^2} k_{22} \left( \frac{1}{2} G_1 - G_{24} \right) + \frac{1}{2 S^2} k_{33} \left(\frac{1}{6} G_1 + \frac{1}{2 S} G_3\right)\nonumber\\
    &- \frac{1}{2 S^2} (k_{22} + k_{24}) \left(G_2 - G_{24}\right)
\end{align}
where the energy contributions associated with the different modes of deformation are grouped analogously to Equation~\ref{eqn:frank}.

\begin{figure}[h!]
	\begin{subfigure}[b]{\linewidth}
		\centering
		\includegraphics[width=\linewidth]{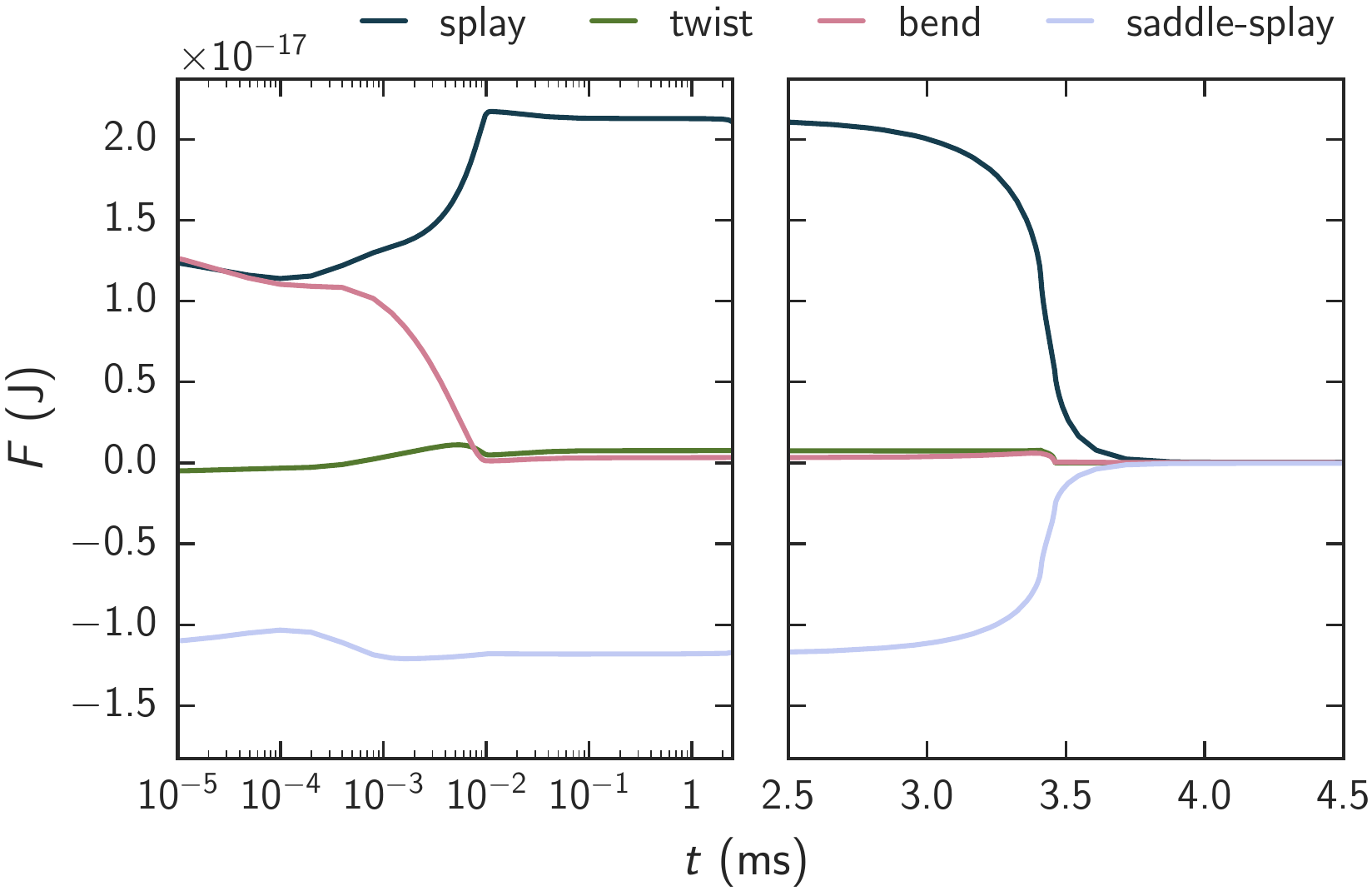}
		\caption{}\label{fig:energy-oneconstant}
	\end{subfigure}
	\begin{subfigure}[b]{\linewidth}
		\centering
		\includegraphics[width=\linewidth]{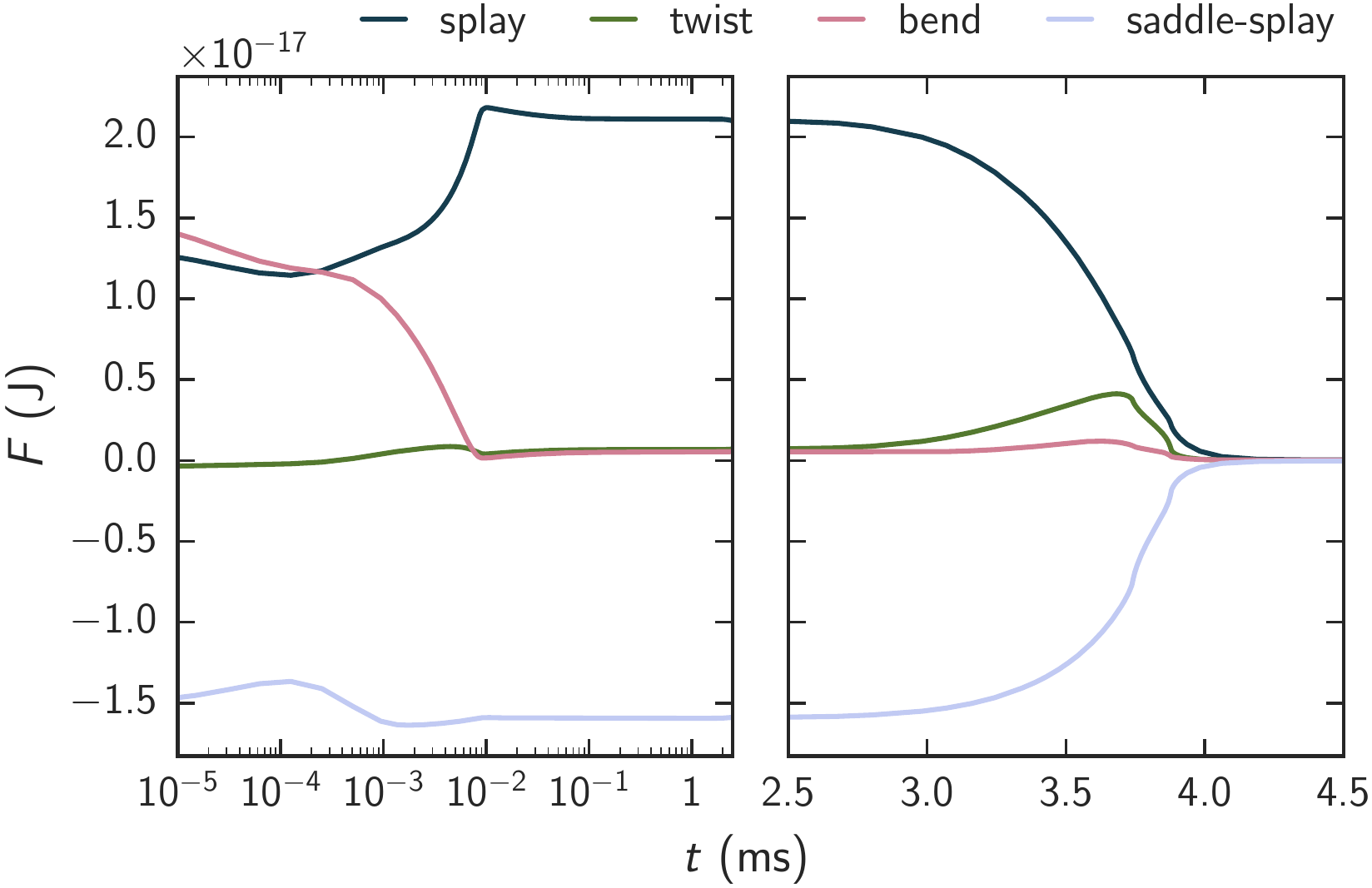}
		\caption{}\label{fig:energy-5cb}
	\end{subfigure}
	\caption{(Color online) Evolution of the approximate contributions of splay, twist, bend, and saddle-splay  deformation modes to the tensorial elastic free energy for the (a) single-constant and (b) 5CB elasticity cases.}
	\label{fig:energy-plot}
\end{figure}

Figure~\ref{fig:energy-plot} shows the evolution of these free energy contributions for both cases of elasticity being studied.
During the nematic formation process, the elastic contributions to the free energy are similar for both cases, in that they are governed by the stable nematic surface layer where surface anchoring promotes a radially-oriented nematic ``shell'' enclosing an unstable isotropic ``core''.
Following nematic formation, their deformation modes predominantly involve high splay deformation, as expected for a radial-like texture.
For the single-constant case (Figure~\ref{fig:energy-oneconstant}), the subsequent radial-to-uniform relaxation process involves a monotonic decrease in the splay deformation contribution until the elasticity vanishes entirely, at which point the domain is uniform.
However, during the radial-to-uniform relaxation process for the 5CB elasticity case (Figure~\ref{fig:energy-5cb}), the splay deformation component decreases more slowly in comparison and is offset by a lesser (energetic) increase in the twist deformation contribution.
This increase in the twist deformation contribution coincides with the twisting relaxation process, which is then followed by a decrease in all elastic contributions as the defect loop escapes.
Intuitively, this result reflects the fact that the twist elastic constant is lower than both splay and bend for 5CB.
Thus, for the case of 5CB elasticity, the splay-dominated relaxation mechanism observed in the single-constant case is a higher-energy dynamic pathway to equilibrium compared to the chiral symmetry-breaking twist mechanism.
This interpretation is synonymous with that for equilibrium twisted-radial  droplet textures where the magnitude of the twist angle in the texture was dependent on the ratio of twist to splay elastic constants \cite{Press1974}.

\begin{figure}[h!]
	\includegraphics[width=\linewidth]{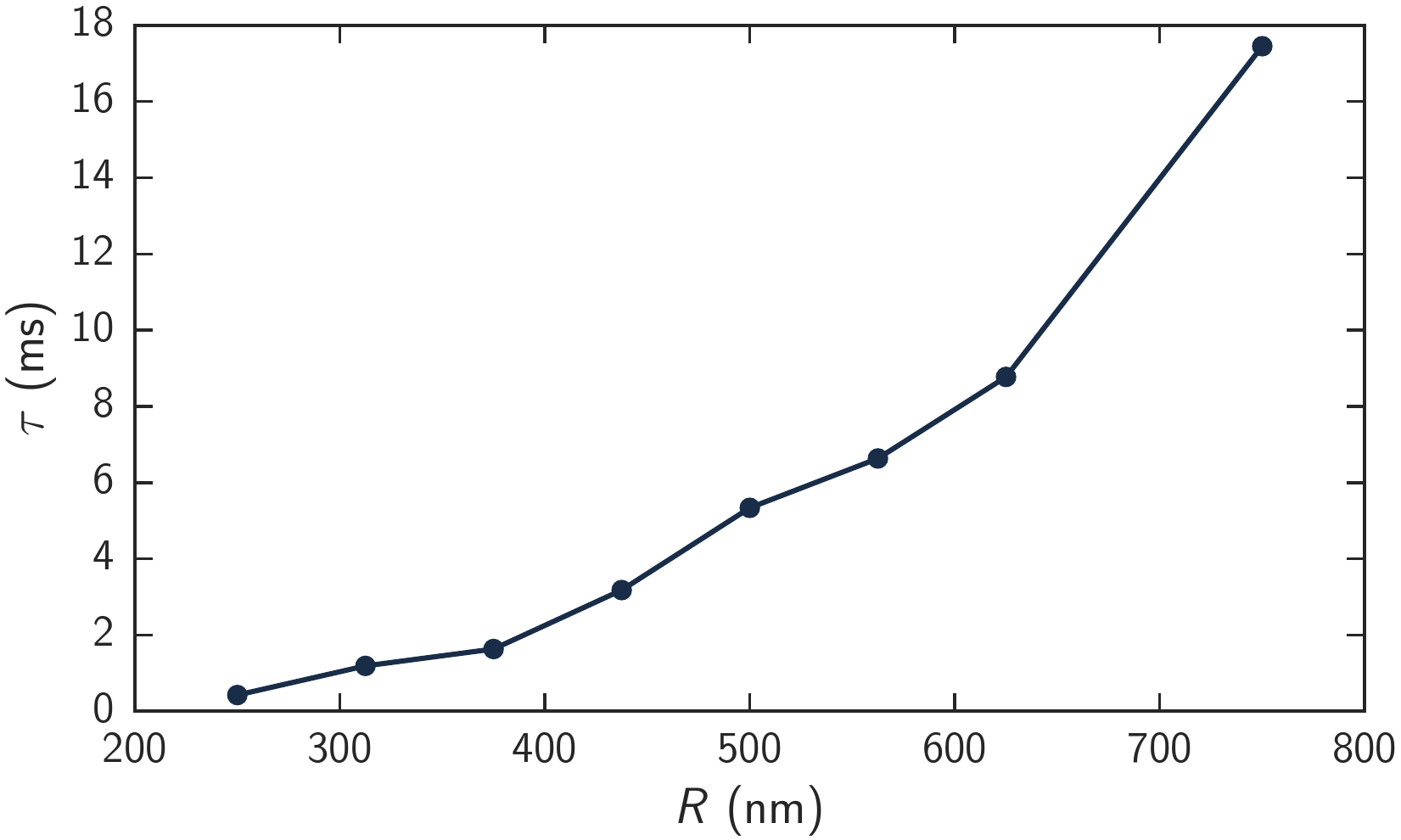}
	\caption{Time scale of the twist relaxation mechanism from radial to uniform nematic texture for the 5CB elasticity case versus droplet radius $R$, measured from the end of the formation process until the escape of the defect loop.}
	\label{fig:time-scales}
\end{figure}

Additional simulations for nematic droplets with radii ranging from $R=250\text{--}\SI{750}{\nano\meter}$ were then performed using the 5CB elasticity parameters, all within a scale and parameter range where the uniform texture is stable.
In all simulations the twist relaxation mechanism was observed and the time scales of the relaxation, which correspond to the region of nonzero twist deformation, were measured (Figure~\ref{fig:time-scales}).
For larger radii (e.g., comparable to those accessible via polarized optical microscopy), it is expected that the radial texture resulting from the nematic formation process would be stable due to the eventual dominance of the surface anchoring energy over the bulk elasticity with increasing droplet size \cite{Williams1986a,Lavrentovich1998}, and thus the twist relaxation mechanism would no longer occur.

In conclusion, continuum simulations of the phase transformation of achiral nematic droplets with weak homeotropic anchoring were found to predict a chiral symmetry-breaking dynamic mechanism in the relaxation of the droplet to a stable uniform texture.
Additionally, it was found that the use of full anisotropy in nematic elasticity, corresponding to 5CB in the present simulations, is required to observe the twisting relaxation mechanism.
In this case, as is true with the class of LC-exhibiting cyanobiphenyl compounds \cite{Madhusudana1982}, twist deformations are energetically favorable compared to splay and bend.
However, a more complex relationship between the nematic elastic constants governing this dynamic mechanism could exist as was found in ref. \cite{Williams1986a} for equilibrium textures.
Finally, the frequently used single-constant elasticity approximation, in which splay, twist, and bend elasticity parameters are assumed equal, is insufficient to predict this relaxation mechanism.

\begin{acknowledgments}
	This work was supported by the Natural Sciences and Engineering Research Council of Canada, Compute Canada, and the Ontario Graduate Scholarship program.
\end{acknowledgments}

\bibliography{self_assembly,pdlcs,computational}

\begin{thebibliography}{33}%
\makeatletter
\providecommand \@ifxundefined [1]{%
 \@ifx{#1\undefined}
}%
\providecommand \@ifnum [1]{%
 \ifnum #1\expandafter \@firstoftwo
 \else \expandafter \@secondoftwo
 \fi
}%
\providecommand \@ifx [1]{%
 \ifx #1\expandafter \@firstoftwo
 \else \expandafter \@secondoftwo
 \fi
}%
\providecommand \natexlab [1]{#1}%
\providecommand \enquote  [1]{``#1''}%
\providecommand \bibnamefont  [1]{#1}%
\providecommand \bibfnamefont [1]{#1}%
\providecommand \citenamefont [1]{#1}%
\providecommand \href@noop [0]{\@secondoftwo}%
\providecommand \href [0]{\begingroup \@sanitize@url \@href}%
\providecommand \@href[1]{\@@startlink{#1}\@@href}%
\providecommand \@@href[1]{\endgroup#1\@@endlink}%
\providecommand \@sanitize@url [0]{\catcode `\\12\catcode `\$12\catcode
  `\&12\catcode `\#12\catcode `\^12\catcode `\_12\catcode `\%12\relax}%
\providecommand \@@startlink[1]{}%
\providecommand \@@endlink[0]{}%
\providecommand \url  [0]{\begingroup\@sanitize@url \@url }%
\providecommand \@url [1]{\endgroup\@href {#1}{\urlprefix }}%
\providecommand \urlprefix  [0]{URL }%
\providecommand \Eprint [0]{\href }%
\providecommand \doibase [0]{http://dx.doi.org/}%
\providecommand \selectlanguage [0]{\@gobble}%
\providecommand \bibinfo  [0]{\@secondoftwo}%
\providecommand \bibfield  [0]{\@secondoftwo}%
\providecommand \translation [1]{[#1]}%
\providecommand \BibitemOpen [0]{}%
\providecommand \bibitemStop [0]{}%
\providecommand \bibitemNoStop [0]{.\EOS\space}%
\providecommand \EOS [0]{\spacefactor3000\relax}%
\providecommand \BibitemShut  [1]{\csname bibitem#1\endcsname}%
\let\auto@bib@innerbib\@empty
\bibitem [{\citenamefont {Volovik}\ and\ \citenamefont
  {Lavrentovich}(1983)}]{Volovik1983}%
  \BibitemOpen
  \bibfield  {author} {\bibinfo {author} {\bibfnamefont {G.~E.}\ \bibnamefont
  {Volovik}}\ and\ \bibinfo {author} {\bibfnamefont {O.~D.}\ \bibnamefont
  {Lavrentovich}},\ }\href@noop {} {\bibfield  {journal} {\bibinfo  {journal}
  {Sov. Phys. JETP}\ }\textbf {\bibinfo {volume} {58}},\ \bibinfo {pages}
  {1159} (\bibinfo {year} {1983})}\BibitemShut {NoStop}%
\bibitem [{\citenamefont {Williams}(1986)}]{Williams1986a}%
  \BibitemOpen
  \bibfield  {author} {\bibinfo {author} {\bibfnamefont {R.~D.}\ \bibnamefont
  {Williams}},\ }\href {\doibase 10.1088/0305-4470/19/16/019} {\bibfield
  {journal} {\bibinfo  {journal} {Journal of Physics A: Mathematical and
  General}\ }\textbf {\bibinfo {volume} {19}},\ \bibinfo {pages} {3211}
  (\bibinfo {year} {1986})}\BibitemShut {NoStop}%
\bibitem [{\citenamefont {Drzaic}(1999)}]{Drzaic1999}%
  \BibitemOpen
  \bibfield  {author} {\bibinfo {author} {\bibfnamefont {P.~S.}\ \bibnamefont
  {Drzaic}},\ }\href {\doibase 10.1080/026782999204660} {\bibfield  {journal}
  {\bibinfo  {journal} {Liquid Crystals}\ }\textbf {\bibinfo {volume} {26}},\
  \bibinfo {pages} {623} (\bibinfo {year} {1999})}\BibitemShut {NoStop}%
\bibitem [{\citenamefont {Ohzono}\ and\ \citenamefont
  {Fukuda}(2012)}]{Ohzono2012}%
  \BibitemOpen
  \bibfield  {author} {\bibinfo {author} {\bibfnamefont {T.}~\bibnamefont
  {Ohzono}}\ and\ \bibinfo {author} {\bibfnamefont {J.-i.}\ \bibnamefont
  {Fukuda}},\ }\href {\doibase 10.1038/ncomms1709} {\bibfield  {journal}
  {\bibinfo  {journal} {Nature Communications}\ }\textbf {\bibinfo {volume}
  {3}},\ \bibinfo {pages} {701} (\bibinfo {year} {2012})}\BibitemShut {NoStop}%
\bibitem [{\citenamefont {Vanzo}\ \emph {et~al.}(2012)\citenamefont {Vanzo},
  \citenamefont {Ricci}, \citenamefont {Berardi},\ and\ \citenamefont
  {Zannoni}}]{Vanzo2012}%
  \BibitemOpen
  \bibfield  {author} {\bibinfo {author} {\bibfnamefont {D.}~\bibnamefont
  {Vanzo}}, \bibinfo {author} {\bibfnamefont {M.}~\bibnamefont {Ricci}},
  \bibinfo {author} {\bibfnamefont {R.}~\bibnamefont {Berardi}}, \ and\
  \bibinfo {author} {\bibfnamefont {C.}~\bibnamefont {Zannoni}},\ }\href@noop
  {} {\bibfield  {journal} {\bibinfo  {journal} {Soft Matter}\ }\textbf
  {\bibinfo {volume} {8}},\ \bibinfo {pages} {11790} (\bibinfo {year}
  {2012})}\BibitemShut {NoStop}%
\bibitem [{\citenamefont {Press}\ and\ \citenamefont
  {Arrott}(1974)}]{Press1974}%
  \BibitemOpen
  \bibfield  {author} {\bibinfo {author} {\bibfnamefont {M.~J.}\ \bibnamefont
  {Press}}\ and\ \bibinfo {author} {\bibfnamefont {A.~S.}\ \bibnamefont
  {Arrott}},\ }\href {\doibase 10.1103/PhysRevLett.33.403} {\bibfield
  {journal} {\bibinfo  {journal} {Physical Review Letters}\ }\textbf {\bibinfo
  {volume} {33}},\ \bibinfo {pages} {403} (\bibinfo {year} {1974})}\BibitemShut
  {NoStop}%
\bibitem [{\citenamefont {Jeong}\ \emph {et~al.}(2014)\citenamefont {Jeong},
  \citenamefont {Davidson}, \citenamefont {Collings}, \citenamefont
  {Lubensky},\ and\ \citenamefont {Yodh}}]{Jeong2014}%
  \BibitemOpen
  \bibfield  {author} {\bibinfo {author} {\bibfnamefont {J.}~\bibnamefont
  {Jeong}}, \bibinfo {author} {\bibfnamefont {Z.~S.}\ \bibnamefont {Davidson}},
  \bibinfo {author} {\bibfnamefont {P.~J.}\ \bibnamefont {Collings}}, \bibinfo
  {author} {\bibfnamefont {T.~C.}\ \bibnamefont {Lubensky}}, \ and\ \bibinfo
  {author} {\bibfnamefont {A.~G.}\ \bibnamefont {Yodh}},\ }\href {\doibase
  10.1073/pnas.1315121111} {\bibfield  {journal} {\bibinfo  {journal}
  {Proceedings of the National Academy of Sciences of the United States of
  America}\ }\textbf {\bibinfo {volume} {111}},\ \bibinfo {pages} {1742}
  (\bibinfo {year} {2014})},\ \Eprint {http://arxiv.org/abs/arXiv:1408.1149}
  {arXiv:arXiv:1408.1149} \BibitemShut {NoStop}%
\bibitem [{\citenamefont {Lin}\ \emph {et~al.}(2011)\citenamefont {Lin},
  \citenamefont {Miller}, \citenamefont {Bertics}, \citenamefont {Murphy},
  \citenamefont {Pablo},\ and\ \citenamefont {Abbott}}]{Lin2011}%
  \BibitemOpen
  \bibfield  {author} {\bibinfo {author} {\bibfnamefont {I.-H.}\ \bibnamefont
  {Lin}}, \bibinfo {author} {\bibfnamefont {D.~S.}\ \bibnamefont {Miller}},
  \bibinfo {author} {\bibfnamefont {P.~J.}\ \bibnamefont {Bertics}}, \bibinfo
  {author} {\bibfnamefont {C.~J.}\ \bibnamefont {Murphy}}, \bibinfo {author}
  {\bibfnamefont {J.~J.~D.}\ \bibnamefont {Pablo}}, \ and\ \bibinfo {author}
  {\bibfnamefont {N.~L.}\ \bibnamefont {Abbott}},\ }\href@noop {} {\bibfield
  {journal} {\bibinfo  {journal} {Science}\ }\textbf {\bibinfo {volume}
  {332}},\ \bibinfo {pages} {1297} (\bibinfo {year} {2011})}\BibitemShut
  {NoStop}%
\bibitem [{\citenamefont {Drzaic}(1995)}]{Drzaic1995}%
  \BibitemOpen
  \bibfield  {author} {\bibinfo {author} {\bibfnamefont {P.~S.}\ \bibnamefont
  {Drzaic}},\ }\href@noop {} {\emph {\bibinfo {title} {Liquid Crystal
  Dispersions}}},\ \bibinfo {series} {Liquid Crystals Series}, Vol.~\bibinfo
  {volume} {1}\ (\bibinfo  {publisher} {World Scientific},\ \bibinfo {year}
  {1995})\BibitemShut {NoStop}%
\bibitem [{\citenamefont {Lavrentovich}(2003)}]{Lavrentovich2003}%
  \BibitemOpen
  \bibfield  {author} {\bibinfo {author} {\bibfnamefont {O.~D.}\ \bibnamefont
  {Lavrentovich}},\ }in\ \href {\doibase 10.1007/978-94-007-1029-0_6} {\emph
  {\bibinfo {booktitle} {Patterns of Symmetry Breaking}}},\ Vol.~\bibinfo
  {volume} {52},\ \bibinfo {editor} {edited by\ \bibinfo {editor}
  {\bibfnamefont {O.~D.}\ \bibnamefont {Lavrentovich}}, \bibinfo {editor}
  {\bibfnamefont {P.}~\bibnamefont {Pasini}}, \bibinfo {editor} {\bibfnamefont
  {C.}~\bibnamefont {Zannoni}}, \ and\ \bibinfo {editor} {\bibfnamefont
  {S.}~\bibnamefont {{\v{Z}}umer}}}\ (\bibinfo  {publisher} {Springer
  Netherlands},\ \bibinfo {address} {Dordrecht},\ \bibinfo {year} {2003})\ pp.\
  \bibinfo {pages} {161--195}\BibitemShut {NoStop}%
\bibitem [{\citenamefont {Crawford}\ \emph {et~al.}(1992)\citenamefont
  {Crawford}, \citenamefont {Allender},\ and\ \citenamefont
  {Doane}}]{Crawford1992}%
  \BibitemOpen
  \bibfield  {author} {\bibinfo {author} {\bibfnamefont {G.}~\bibnamefont
  {Crawford}}, \bibinfo {author} {\bibfnamefont {D.}~\bibnamefont {Allender}},
  \ and\ \bibinfo {author} {\bibfnamefont {J.}~\bibnamefont {Doane}},\ }\href
  {\doibase 10.1103/PhysRevA.45.8693} {\bibfield  {journal} {\bibinfo
  {journal} {Physical Review A}\ }\textbf {\bibinfo {volume} {45}},\ \bibinfo
  {pages} {8693} (\bibinfo {year} {1992})}\BibitemShut {NoStop}%
\bibitem [{\citenamefont {Higgins}\ \emph {et~al.}(2005)\citenamefont
  {Higgins}, \citenamefont {Hall},\ and\ \citenamefont {Xie}}]{Higgins2005}%
  \BibitemOpen
  \bibfield  {author} {\bibinfo {author} {\bibfnamefont {D.~A.}\ \bibnamefont
  {Higgins}}, \bibinfo {author} {\bibfnamefont {J.~E.}\ \bibnamefont {Hall}}, \
  and\ \bibinfo {author} {\bibfnamefont {A.}~\bibnamefont {Xie}},\ }\href
  {\doibase 10.1021/ar040106p} {\bibfield  {journal} {\bibinfo  {journal}
  {Accounts of chemical research}\ }\textbf {\bibinfo {volume} {38}},\ \bibinfo
  {pages} {137} (\bibinfo {year} {2005})}\BibitemShut {NoStop}%
\bibitem [{\citenamefont {Abukhdeir}(2016)}]{Abukhdeir2016a}%
  \BibitemOpen
  \bibfield  {author} {\bibinfo {author} {\bibfnamefont {N.~M.}\ \bibnamefont
  {Abukhdeir}},\ }\href {\doibase 10.1080/02678292.2016.1239772} {\bibfield
  {journal} {\bibinfo  {journal} {Liquid Crystals}\ }\textbf {\bibinfo {volume}
  {43}},\ \bibinfo {pages} {2300} (\bibinfo {year} {2016})}\BibitemShut
  {NoStop}%
\bibitem [{\citenamefont {Lavrentovich}(1998)}]{Lavrentovich1998}%
  \BibitemOpen
  \bibfield  {author} {\bibinfo {author} {\bibfnamefont {O.~D.}\ \bibnamefont
  {Lavrentovich}},\ }\href {\doibase 10.1080/026782998207640} {\bibfield
  {journal} {\bibinfo  {journal} {Liquid Crystals}\ }\textbf {\bibinfo {volume}
  {24}},\ \bibinfo {pages} {117} (\bibinfo {year} {1998})}\BibitemShut
  {NoStop}%
\bibitem [{\citenamefont {de~Gennes}\ and\ \citenamefont
  {Prost}(1995)}]{deGennes1995}%
  \BibitemOpen
  \bibfield  {author} {\bibinfo {author} {\bibfnamefont {P.}~\bibnamefont
  {de~Gennes}}\ and\ \bibinfo {author} {\bibfnamefont {J.}~\bibnamefont
  {Prost}},\ }\href@noop {} {\emph {\bibinfo {title} {The Physics of Liquid
  Crystals}}},\ \bibinfo {edition} {2nd}\ ed.\ (\bibinfo  {publisher} {Oxford
  University Press},\ \bibinfo {address} {New York},\ \bibinfo {year}
  {1995})\BibitemShut {NoStop}%
\bibitem [{\citenamefont {Sonnet}\ \emph {et~al.}(1995)\citenamefont {Sonnet},
  \citenamefont {Kilian},\ and\ \citenamefont {Hess}}]{Sonnet1995}%
  \BibitemOpen
  \bibfield  {author} {\bibinfo {author} {\bibfnamefont {A.}~\bibnamefont
  {Sonnet}}, \bibinfo {author} {\bibfnamefont {A.}~\bibnamefont {Kilian}}, \
  and\ \bibinfo {author} {\bibfnamefont {S.}~\bibnamefont {Hess}},\ }\href
  {\doibase 10.1103/PhysRevE.52.718} {\bibfield  {journal} {\bibinfo  {journal}
  {Phys. Rev. E}\ }\textbf {\bibinfo {volume} {52}},\ \bibinfo {pages} {718}
  (\bibinfo {year} {1995})}\BibitemShut {NoStop}%
\bibitem [{\citenamefont {Barbero}\ and\ \citenamefont
  {Evangelista}(2001)}]{Barbero2001}%
  \BibitemOpen
  \bibfield  {author} {\bibinfo {author} {\bibfnamefont {G.}~\bibnamefont
  {Barbero}}\ and\ \bibinfo {author} {\bibfnamefont {L.~R.}\ \bibnamefont
  {Evangelista}},\ }\href@noop {} {\emph {\bibinfo {title} {{An Elementary
  Course on the Continuum Theory for Nematic Liquid Crystals}}}}\ (\bibinfo
  {publisher} {World Scientific},\ \bibinfo {year} {2001})\BibitemShut
  {NoStop}%
\bibitem [{\citenamefont {Mori}\ \emph {et~al.}(1999)\citenamefont {Mori},
  \citenamefont {Gartland}, \citenamefont {Kelly},\ and\ \citenamefont
  {Bos}}]{Mori1999}%
  \BibitemOpen
  \bibfield  {author} {\bibinfo {author} {\bibfnamefont {H.}~\bibnamefont
  {Mori}}, \bibinfo {author} {\bibfnamefont {E.~C.}\ \bibnamefont {Gartland}},
  \bibinfo {author} {\bibfnamefont {J.~R.}\ \bibnamefont {Kelly}}, \ and\
  \bibinfo {author} {\bibfnamefont {P.~J.}\ \bibnamefont {Bos}},\ }\href
  {\doibase 10.1143/JJAP.38.135} {\bibfield  {journal} {\bibinfo  {journal}
  {Japanese Journal of Applied Physics, Part 1: Regular Papers and Short Notes
  and Review Papers}\ }\textbf {\bibinfo {volume} {38}},\ \bibinfo {pages}
  {135} (\bibinfo {year} {1999})}\BibitemShut {NoStop}%
\bibitem [{\citenamefont {Fu}\ and\ \citenamefont {Abukhdeir}(2017)}]{Fu2017}%
  \BibitemOpen
  \bibfield  {author} {\bibinfo {author} {\bibfnamefont {F.}~\bibnamefont
  {Fu}}\ and\ \bibinfo {author} {\bibfnamefont {N.~M.}\ \bibnamefont
  {Abukhdeir}},\ }\href {\doibase 10.1039/C7SM00484B} {\bibfield  {journal}
  {\bibinfo  {journal} {Soft Matter}\ } (\bibinfo {year} {2017}),\
  10.1039/C7SM00484B}\BibitemShut {NoStop}%
\bibitem [{\citenamefont {Frank}(1958)}]{Frank1958}%
  \BibitemOpen
  \bibfield  {author} {\bibinfo {author} {\bibfnamefont {F.}~\bibnamefont
  {Frank}},\ }\href@noop {} {\bibfield  {journal} {\bibinfo  {journal}
  {Discussions of the Faraday Society}\ }\textbf {\bibinfo {volume} {25}},\
  \bibinfo {pages} {19} (\bibinfo {year} {1958})}\BibitemShut {NoStop}%
\bibitem [{\citenamefont {Bogi}\ and\ \citenamefont {Faetti}(2001)}]{Bogi2001}%
  \BibitemOpen
  \bibfield  {author} {\bibinfo {author} {\bibfnamefont {A.}~\bibnamefont
  {Bogi}}\ and\ \bibinfo {author} {\bibfnamefont {S.}~\bibnamefont {Faetti}},\
  }\href {\doibase 10.1080/02678290010021589} {\bibfield  {journal} {\bibinfo
  {journal} {Liquid Crystals}\ }\textbf {\bibinfo {volume} {28}},\ \bibinfo
  {pages} {729} (\bibinfo {year} {2001})}\BibitemShut {NoStop}%
\bibitem [{\citenamefont {Dunmur}\ \emph {et~al.}(2001)\citenamefont {Dunmur},
  \citenamefont {Fukuda},\ and\ \citenamefont {Luckhurst}}]{Dunmur2001}%
  \BibitemOpen
  \bibfield  {author} {\bibinfo {author} {\bibfnamefont {D.~A.}\ \bibnamefont
  {Dunmur}}, \bibinfo {author} {\bibfnamefont {A.}~\bibnamefont {Fukuda}}, \
  and\ \bibinfo {author} {\bibfnamefont {G.}~\bibnamefont {Luckhurst}},\
  }\href@noop {} {\emph {\bibinfo {title} {{Physical Properties of Liquid
  Crystals: Nematics}}}}\ (\bibinfo  {publisher} {Institution of Engineering
  and Technology},\ \bibinfo {year} {2001})\BibitemShut {NoStop}%
\bibitem [{\citenamefont {Polak}\ \emph {et~al.}(1994)\citenamefont {Polak},
  \citenamefont {Crawford}, \citenamefont {Kostival}, \citenamefont {Doane},\
  and\ \citenamefont {Zumer}}]{Polak1994}%
  \BibitemOpen
  \bibfield  {author} {\bibinfo {author} {\bibfnamefont {R.~D.}\ \bibnamefont
  {Polak}}, \bibinfo {author} {\bibfnamefont {G.~P.}\ \bibnamefont {Crawford}},
  \bibinfo {author} {\bibfnamefont {B.~C.}\ \bibnamefont {Kostival}}, \bibinfo
  {author} {\bibfnamefont {J.~W.}\ \bibnamefont {Doane}}, \ and\ \bibinfo
  {author} {\bibfnamefont {S.}~\bibnamefont {Zumer}},\ }\href {\doibase
  10.1103/PhysRevE.49.R978} {\bibfield  {journal} {\bibinfo  {journal}
  {Physical Review E}\ }\textbf {\bibinfo {volume} {49}},\ \bibinfo {pages}
  {R978} (\bibinfo {year} {1994})}\BibitemShut {NoStop}%
\bibitem [{\citenamefont {Pairam}\ \emph {et~al.}(2013)\citenamefont {Pairam},
  \citenamefont {Vallamkondu}, \citenamefont {Koning}, \citenamefont {van
  Zuiden}, \citenamefont {Ellis}, \citenamefont {Bates}, \citenamefont
  {Vitelli},\ and\ \citenamefont {Fernandez-Nieves}}]{Pairam2013}%
  \BibitemOpen
  \bibfield  {author} {\bibinfo {author} {\bibfnamefont {E.}~\bibnamefont
  {Pairam}}, \bibinfo {author} {\bibfnamefont {J.}~\bibnamefont {Vallamkondu}},
  \bibinfo {author} {\bibfnamefont {V.}~\bibnamefont {Koning}}, \bibinfo
  {author} {\bibfnamefont {B.~C.}\ \bibnamefont {van Zuiden}}, \bibinfo
  {author} {\bibfnamefont {P.~W.}\ \bibnamefont {Ellis}}, \bibinfo {author}
  {\bibfnamefont {M.~A.}\ \bibnamefont {Bates}}, \bibinfo {author}
  {\bibfnamefont {V.}~\bibnamefont {Vitelli}}, \ and\ \bibinfo {author}
  {\bibfnamefont {A.}~\bibnamefont {Fernandez-Nieves}},\ }\href {\doibase
  10.1073/pnas.1221380110} {\bibfield  {journal} {\bibinfo  {journal}
  {Proceedings of the National Academy of Sciences of the United States of
  America}\ }\textbf {\bibinfo {volume} {110}},\ \bibinfo {pages} {9295}
  (\bibinfo {year} {2013})},\ \Eprint {http://arxiv.org/abs/1212.1771}
  {arXiv:1212.1771} \BibitemShut {NoStop}%
\bibitem [{\citenamefont {Nobili}\ and\ \citenamefont
  {Durand}(1992)}]{Nobili1992}%
  \BibitemOpen
  \bibfield  {author} {\bibinfo {author} {\bibfnamefont {M.}~\bibnamefont
  {Nobili}}\ and\ \bibinfo {author} {\bibfnamefont {G.}~\bibnamefont
  {Durand}},\ }\href {\doibase 10.1103/PhysRevA.46.R6174} {\bibfield  {journal}
  {\bibinfo  {journal} {Physical Review A}\ }\textbf {\bibinfo {volume} {46}},\
  \bibinfo {pages} {R6174} (\bibinfo {year} {1992})}\BibitemShut {NoStop}%
\bibitem [{\citenamefont {Hohenberg}\ and\ \citenamefont
  {Halperin}(1977)}]{Hohenberg1977}%
  \BibitemOpen
  \bibfield  {author} {\bibinfo {author} {\bibfnamefont {P.~C.}\ \bibnamefont
  {Hohenberg}}\ and\ \bibinfo {author} {\bibfnamefont {B.~I.}\ \bibnamefont
  {Halperin}},\ }\href {\doibase 10.1103/RevModPhys.49.435} {\bibfield
  {journal} {\bibinfo  {journal} {Rev. Mod. Phys.}\ }\textbf {\bibinfo {volume}
  {49}},\ \bibinfo {pages} {435} (\bibinfo {year} {1977})}\BibitemShut
  {NoStop}%
\bibitem [{\citenamefont {Skarp}\ \emph {et~al.}(1980)\citenamefont {Skarp},
  \citenamefont {Lagerwall},\ and\ \citenamefont {Stebler}}]{Skarp1980}%
  \BibitemOpen
  \bibfield  {author} {\bibinfo {author} {\bibfnamefont {K.}~\bibnamefont
  {Skarp}}, \bibinfo {author} {\bibfnamefont {S.}~\bibnamefont {Lagerwall}}, \
  and\ \bibinfo {author} {\bibfnamefont {B.}~\bibnamefont {Stebler}},\
  }\href@noop {} {\bibfield  {journal} {\bibinfo  {journal} {Molecular Crystals
  and Liquid Crystals}\ }\textbf {\bibinfo {volume} {60}},\ \bibinfo {pages}
  {215} (\bibinfo {year} {1980})}\BibitemShut {NoStop}%
\bibitem [{\citenamefont {Sheng}(1982)}]{Sheng1982}%
  \BibitemOpen
  \bibfield  {author} {\bibinfo {author} {\bibfnamefont {P.}~\bibnamefont
  {Sheng}},\ }\href {\doibase 10.1103/PhysRevA.26.1610} {\bibfield  {journal}
  {\bibinfo  {journal} {Phys. Rev. A}\ }\textbf {\bibinfo {volume} {26}},\
  \bibinfo {pages} {1610} (\bibinfo {year} {1982})}\BibitemShut {NoStop}%
\bibitem [{\citenamefont {Logg}\ \emph {et~al.}(2012)\citenamefont {Logg},
  \citenamefont {Mardal},\ and\ \citenamefont {Wells}}]{FEniCSBook}%
  \BibitemOpen
  \bibfield  {author} {\bibinfo {author} {\bibfnamefont {A.}~\bibnamefont
  {Logg}}, \bibinfo {author} {\bibfnamefont {K.-A.}\ \bibnamefont {Mardal}}, \
  and\ \bibinfo {author} {\bibfnamefont {G.}~\bibnamefont {Wells}},\
  }\href@noop {} {\emph {\bibinfo {title} {Automated solution of differential
  equations by the finite element method: The {FEniCS} book}}},\ Vol.~\bibinfo
  {volume} {84}\ (\bibinfo  {publisher} {Springer Science \& Business Media},\
  \bibinfo {year} {2012})\BibitemShut {NoStop}%
\bibitem [{\citenamefont {Khayyatzadeh}(2014)}]{Khayyatzadeh2014}%
  \BibitemOpen
  \bibfield  {author} {\bibinfo {author} {\bibfnamefont {P.}~\bibnamefont
  {Khayyatzadeh}},\ }\emph {\bibinfo {title} {{Geometry and Anchoring Effects
  on Elliptic Cylinder Domains of Nematic Phases}}},\ \href@noop {} {Master's
  thesis},\ \bibinfo  {school} {University of Waterloo} (\bibinfo {year}
  {2014})\BibitemShut {NoStop}%
\bibitem [{\citenamefont {Delmarcelle}\ and\ \citenamefont
  {Hesselink}(1993)}]{Delmarcelle1993}%
  \BibitemOpen
  \bibfield  {author} {\bibinfo {author} {\bibfnamefont {T.}~\bibnamefont
  {Delmarcelle}}\ and\ \bibinfo {author} {\bibfnamefont {L.}~\bibnamefont
  {Hesselink}},\ }\href {\doibase
  http://doi.ieeecomputersociety.org/10.1109/38.219447} {\bibfield  {journal}
  {\bibinfo  {journal} {IEEE Comput. Graph.}\ }\textbf {\bibinfo {volume}
  {13}},\ \bibinfo {pages} {25} (\bibinfo {year} {1993})}\BibitemShut {NoStop}%
\bibitem [{\citenamefont {Fu}\ and\ \citenamefont {Abukhdeir}(2015)}]{Fu2015}%
  \BibitemOpen
  \bibfield  {author} {\bibinfo {author} {\bibfnamefont {F.}~\bibnamefont
  {Fu}}\ and\ \bibinfo {author} {\bibfnamefont {N.}~\bibnamefont {Abukhdeir}},\
  }\href {\doibase 10.1109/TVCG.2014.2363828} {\bibfield  {journal} {\bibinfo
  {journal} {Visualization and Computer Graphics, IEEE Transactions on}\
  }\textbf {\bibinfo {volume} {21}},\ \bibinfo {pages} {413} (\bibinfo {year}
  {2015})}\BibitemShut {NoStop}%
\bibitem [{\citenamefont {Madhusudana}\ and\ \citenamefont
  {Pratibha}(1982)}]{Madhusudana1982}%
  \BibitemOpen
  \bibfield  {author} {\bibinfo {author} {\bibfnamefont {N.~V.}\ \bibnamefont
  {Madhusudana}}\ and\ \bibinfo {author} {\bibfnamefont {R.}~\bibnamefont
  {Pratibha}},\ }\href {\doibase 10.1080/00268948208074481} {\bibfield
  {journal} {\bibinfo  {journal} {Molecular Crystals and Liquid Crystals}\
  }\textbf {\bibinfo {volume} {89}},\ \bibinfo {pages} {249} (\bibinfo {year}
  {1982})}\BibitemShut {NoStop}%
\end{thebibliography}%

\end{document}